# Profiling Internet Users' Participation in Social Change Agendas: An Application of Q-Methodology

*Josephine Previte, Greg Hearn and Susan Dann*

Faculty of Business, Queensland University of Technology

Brisbane, Australia

## Abstract

New computer-mediated channels of communication are action oriented and have the ability to deliver information and dialogue—moderated and unmoderated—which can facilitate the bringing together society's stakeholders, opinion leaders and change agents, who have the ability to influence social action.  However, existing online studies have been limited in explaining Internet users' willingness to participate in social change agendas online.   They have relied predominately on basic demographic descriptors such as age, education, income and access to technology and have ignored, social, psychological and attitudinal variables that may explain online participation and social change.

The emphasis on social/attitudinal profiling of users is suggested because Internet-based technologies have unique characteristics which impose/reconstruct individual behaviour and participation in online relationships.  Issues of anonymity, perceived risk and trust all influence online behaviour and wiliness to participate in online social change agendas.

Using a social marketing framework combined with social network theory, this study uses Q-methodology to profile the roles and relationships of various stakeholders involved in diffusing social ideas and actions in an online context.  The empirical findings from the research have informed the development of five idealized models based on Internet users' beliefs and attitudes about social change programs in online environments.  In conclusion we make a number of suggestions about devising online social change strategies that engage the diversity of Internet users, and suggest that future research continue to profile the dynamic and complex nature of online users using a Q method approach.

## Introduction

The Internet[1] is a network technology that has the potential to be more persuasive and effective in diffusing social ideas and actions within a global community of interest than any other communication technology in history. What is of interest to social marketers and other social scientists is the development of computer-mediated networks that connect social change advocates with policymakers and other upstream gatekeepers who effect change at a wider community level in society. But as social scientists do we know enough about the people in the government, community and business sectors who are committed to connecting in new ways together online? And do we understand what motivates people to initiate and become involved in these new opportunities for human intervention? In this paper we argue that the majority of existing studies of Internet users provide only limited insights into what motivates people to participate and commit to online social action. We then introduce Q methodology as an alternative research approach for exploring the common attitudinal structures of Internet users that influence their participation in online social change agendas.

## A Social Marketing Framework

The adoption of the Internet as a commercial and social technology by increasing segments of consumer society presents significant opportunities for future planning and implementation of social marketing campaigns. These opportunities are founded on the nature of the Internet as a social technology (Sproull and Faraja 1997) that allows people to gather in communities of interest and sustain connections over time. For those who are unfamiliar with social marketing, the following section briefly outlines its definition and framework. We then discuss the important themes from the Internet sociology literature that inform our framework and establish a case for bringing a social network approach to the social marketing domain to better examine the role of the Internet in social change management strategies.

---

[1] The Internet refers to Websites, listservs, discussion lists and email.

## *What is social marketing?*

The fundamental purpose of social marketing programs is to promote change (Young 2000). Specifically, social marketing "addresses behaviours of social concern, with implications for the social fabric within community" (Slater 2000: 12). In general terms, social marketing is effective when it results in positive behaviour change at a broader societal level. Andreasen's (1995) definition of social marketing is most widely accepted as it contains a number of features which clearly sets the social marketing approach apart from other social change approaches such as the education approach, the persuasion approach, the behavioural modification approach and the social influence approach. Andreasen's formal definition states:

> Social marketing is the application of commercial marketing technologies to the analysis, planning, execution, and evaluation of programs designed to influence the voluntary behaviour of target audiences in order to improve their personal welfare and that of their society.
>
> (Andreasen 1995: 7)

Much of social marketing focuses on individual behaviour change. For example, communication campaigns in social marketing focus on influencing individual behaviour so that a person will take the advocated path towards better health and better living (Goldberg 1995). The focus on individual behaviour occurs because social marketers plan positive behaviour change by relying on consumer behaviour theories which "assume that choices are made based on a calculation of the ratios of personal benefits to personal costs for various courses of action" (Andreasen 1997: 193). But, as Goldberg highlights (1995: 357) "research focusing on the individual is inherently conservative in that it implicitly endorses the status quo with regard to factors associated with the social structure." Considering this point, he argues for the inclusion of "upstream marketing and research." That is, to focus not just on the individual (downstream) behaviours, but to also consider the important influence of public policy issues and the social context of the problems being addressed as critical elements in the overall development of social marketing strategies.

The study of the Internet and its network communities will contribute to research and understandings about the role of upstream gatekeepers and their influence on social change agendas. For example, research by Lennie, Grace, Daws and Simpson (1999) demonstrates the benefits of bringing together a series of society's stakeholders to form a 'virtual community'. Significantly, their research about rural women's use of interactive communication technologies, reported that one of the most empowering aspect of the 'virtual community' they studied, was the connection established between rural and isolated women, with women and men in strategic positions in cities (ie. government, academia and business positions). Further, research by Dann and Dann (2001) into cybercommunity highlights the most distinguishing feature of such communities to be that they combine elements of content and communication into value-added functions.

**Internet Sociology**

There are three issues raised in the Internet sociology literature which inform a social marketing. First is the discussion concerning the emergence of "new" and empowered consumers. Second is the explanation of the cultural context of the Internet. Third is the work highlighting the emergence of community in online environments.

Prevalent across the Internet sociology literature is a line of reasoning which states that "old" consumers have transformed into new, powerful consumers based on their easy access to countless information sources, a vast range of products and services, and greatly extended communities. This has put new consumers in an unprecedented position of control. The new consumer knows more, has more choices, and can act with fewer logistical constraints than ever before (Windham and Orton 2000). This means there are serious implications for those who are *disconnected* from these new intangible resources available through the Internet.

Also central to the consumer issues in this literature is the cultural context of new communication networks. Castells (1998) argues that such networks do not facilitate 'a new culture, in a traditional sense of a system of values, because the multiplicity of subjects in the network and the diversity of networks reject such unifying network culture'. Instead he points out that these new communication networks are 'made of many cultures, many values, many projects, that cross through the minds and inform the strategies of the various participants' (Castells cited in Slevin 2000: 2).

An extensive amount of Internet sociology discusses the development of network relationships and virtual communities. There are critics who take the view that online relationships are impersonal or even hostile, and advocates who perceive it as a way of emancipating interpersonal relationships from physical places (Parks and Floyd cited in Soberg 1999). This argument does not negate the fact that computer-mediated communication brings new challenges to our understanding and perception of social relations.

### Social network theory

Research completed by sociologists such as Wellman and Guila (1999) and Haythornthwaite (2000; 1998) apply social network analysis as a method to explore and evaluate Internet users' participation in online communities. Some of this research has identified that virtual communities can resemble real-life communities in that they give Internet users access to a range of non-material social resources such as emotional support, companionship, information and a sense of belonging (Preece and Ghozati 2001; Wellman and Gulia 1999). Other studies also demonstrate that the Internet's technological and social structure provides support in other ways such as online activism and mobilization of people to participate in social actions (Diani 2000; Ayres 1999). A distinctive feature of the Internet however, is that the provision of these non-material resources are exchanged between individuals who hardly know each other offline or who are total strangers (Wellman and Gulia 1999).

Two other important learnings from the sociology literature, which are informative, are the theory of "strength of weak ties" (Granovetter 1973) and the research on trust. The "strength of weak ties theory" examines the linking and sharing of resources between people and analyses how individual network "ties" influence diffusion and information sharing and community organisation. "Trust"[2] is considered an important element that drives or inhibits online behaviour and participation at both social and commercial levels (Hoffman and Novak 1998). Examining the role of "trust" and the strength of ties between online participants is important to understanding the potential drivers and inhibitors to participation and individual involvement in online social change strategies.

---

[2] The context of trust we are focusing on is *social trust*, not trust in government or other social institutions. As Putnam (2000: 137) highlights "trust in other people is logically quite different from trust in institutions and political authorities. One could easily trust one's neighbour and distrust city hall, or vice versa."

Thus learnings from the social network and the Internet sociology literature extend the social marketing framework guiding this research. Of interest are the range of communication strategies and programs that employ the full spectrum of protest activity from movements, coalitions, networks, protests, petitions and declarations. Such strategies clearly identify the individual's role in behaviour change, as well as the impact of social and political contexts on individual behaviour and the wider community. This proposition is probably contentious to those authors who identify strategies such as online activism with media advocacy and believe that social marketing and media advocacy are opposing approaches to social change[3]. This is a false dichotomy (Smith 2000). For as Slater, Kelly and Edwards (2000: 135) argue, "the integration of media advocacy and social marketing efforts is logical and is not only a tool to support community development around prevention, but may become a principal foci for initiating individuals to participate in community-based interventions." Therefore using a multidisciplinary framework in this research provides unique insights into online behaviours and raises new questions about Internet-based networks that facilitate a social change agenda.

## Profiling Internet Users

There are no precedents for the types of social changes that the Internet is creating each day. One reason for this, is that the Internet is a "quasi-public" space that is open not only to gambling and pornography, but also to e-commerce and to social activities. These social activities include people sharing their feelings, building community networks, or enhancing disadvantaged communities (Mantovani 2001). Ultimately however, it will be the types of services available on the Internet and the amount of disposable time and resources that users have to participate in online activity that will influence and shape the profile of Internet users (Hearn, Mandeville and Anthony 1998).

Predominately Internet users are described by an evolving demographic profile (gender, age, income, education and ethnicity). This is not surprising given that the majority of user profiling has been undertaken to measure potential market share and inform "new" segmentation strategies. Consequently, most of the questions asked about Internet consumers have been, "who's out there?" and "where are they going?" and "what are they doing?" But what if we consider the Internet to be more than a system that facilitates employment and contributes to the economy? What if we think about the Internet as being a network system that can contribute to quality of life, to people's feelings of inclusion, and their sense of empowerment as citizens (Phipps 2000: 62), either locally or globally?

---

[3] See for instance Wallack 1990.

### *The Internet's relational aspects*

Howard Rheingold (1994; 2000) has described intense and lasting relationships in the WELL community, in which members support each other through sickness and even in the face of terminal illness and death. Others, such as Peerce and Ghozait's (2001), have discussed how empathic communication online impacts on individual involvement and participation in communities of interest and also on an individual's sense of well-being. Earlier research by Sproull & Kiesler (1997) provided further insight into the nature of online relationships. They reported that some online relationships are long lasting, but that many are also transient in nature and that in some circumstances, anonymity and not having to meet encouraged openness. They argued that by choosing an obscure login name people could be anonymous, thus protecting them from everyday biases associated with gender, status, race, and age.

In this paper we have engaged with the dialogue surroundings peoples' experiences online that describe human systems, bonds of trust, mutual respect and reciprocity as a part of 'the web of connection' that links individuals into communities and relationships online. We believe that this will better inform an understanding of online exchanges and establish profiles that illustrate the complexity of online interactions.

## An Overview of Q Methodology

Q methodology is a research method that actively explores diverse positions that pervade debates and social discourses on social and economic issues in contemporary life (Brown. Durning and Selden 1999). Q method is especially suited to uncovering the positions held by participants in a debate, rather than accepting the categories developed by either researchers or observers.

As a research technique, Q methodology provides an alternative to typical survey-based research methods, or what is entitled R methodology.[4] Fundamentally the distinction between the two approaches is that an R study focuses on pre-specified independent categories chosen by the researcher(s), whereas Q methodology looks to unravelling the complex subjectivities in people's beliefs, attitudes and behaviours around a specific social issue. Addams and Proops (2000:1) state:

> … in contrast to other research methods, not only does Q systematically identify groups of individuals with common attitude structures by seeking patterns of response across individuals, rather than variables, it also allows individuals to 'speak for themselves' rather than use pre-specified measures.

**The Pilot Study Procedure**

This pilot study applied a typical Q methodology paradigm. Firstly, a number of interviews and focus groups were conducted with Internet users to ground the study in the discourse[5] of Internet users. Data collection via interviews is an important stage of the research process as it focuses the research on issues which are mostly or wholly raised by participants, rather than the researcher. Interview participants were recruited to the study based on their level of experience and differences in opinion expressed towards Internet-based technologies. Participants' levels of experience ranged from Internet users with only limited knowledge and involvement in online exchanges, to those users who identified the Internet as an addictive behaviour. Individual attitudes to the technology ranged from the cynical or pessimistic to the overtly positive and optimistic.

Secondly we selected approximately 150 statements that were representative of the diversity of communication about attitudes towards participation in online relationships. The statements were initially sorted into a seven-cell typology (see Table 1) and then narrowed to the final 50 statements used in the Q-sorts.

---

[4] For a detailed discussion on the completing dialogue between Q and R Methodologies read Brown, S.R. (1993), "A primer on Q methodology", *Operant Subjectivity*, 16, 91-139 and Brown, S.R., D.W. Durning and S.C. Selden (1999), "Q methodology", in G.J. Miller and M.L. Whicker (eds), *Handbook of Research Methods in Public Administration*, New York: Marcel Dekker, pp.599-637.

[5] Discourse is used here to mean a way of seeing and talking about something (Addams 2000)

Next, a further thirty-two participants were selected to complete 'Q-sorts'. A convenience sampling approach was used to recruit research participants. Participants were selected based on their familiarity with different forms of online communication and accessibility to the Internet. Each participant was asked to sort the statements in the sample in a standard Q-sort fashion.[6] In doing so, each participant effectively modelled their individual point of view about participating in social change strategies facilitated by the Internet. Finally the thirty-two Q-sorts were statistically (factor) analysed and the extraction of five 'typical' Q-sorts were obtained. These five profiles represent the distinctive collective understanding of beliefs and attitudes towards online participation in social change strategies. We will describe and discuss these five profiles in the following section.

*Table 1: Statement Design Categories*

| | **Statement design category** | **Statement subgroup** |
|---|---|---|
| A | General attitudes to Internet technology | Impacts of Internet on broader society, Optimistic and pessimistic statements concerning the impact of technology in general |
| B | Internet-based functions and actions | Social action (eg. Petitions); Agenda setting; Active information searching; |
| C | Internet networks | Participation in online groups; Commitment |
| D | Internet relationships | Establishment and maintenance of strong and weak tie relationships online |
| E | Internet communities | Attitude and beliefs towards online community; Involvement in online community |
| F | Internet-based information | Information quality; Types of information (Bulletin Boards, email, chat exchanged) |
| G | Internet characteristics | Enhancing interaction; Anonymity, Trust |

### *The research sample*

A convenience sample of professionals was drawn predominately from universities. To broaden the opinion-base of the sample we used a snowballing technique to include other

---

[6] The standard Q sort fashion is a quasi-normal distribution. For this study the researchers used an opinion continuum from –4 to 4. The extremes of the distribution were coded +4 for 'most agree' and –4 for 'most disagree', with 0 indicating indifference. Each participant was asked to sort 50 statements.

Internet users from industry and social sectors who had high levels of familiarity and accessibility to the Internet. The targeting of professionals was considered appropriate because the "typical" demographic profile of an Internet user is someone who is male, has an above average income, tertiary (college) education training or higher, and aged between 20-30 years. Of the thirty-two participants involved in the pilot study seventeen were women and fifteen were men, and their age ranged from 22 –55 years. When asked to describe their participation and involvement in online activities[7], eleven said they had a moderate level of involvement, fifteen identified with having an intermediate level of involvement and six said they were highly involved in relational and social exchanges online.

The pilot research sample had a larger number of women than men, and was older than the "typical" Internet user profiled. However, we felt it was more important for the pilot sample group's Internet experience not to be limited because of social or economic barriers to access. In fact, our sample represents skilled users of the Internet, who have not experienced any significant barriers to participation in social or economic activities online. More importantly, in sampling Internet users for the research, our focus was on ensuring that the full diversity of *opinion* was present in the discourse of research participants. In Q method this is more important when selecting research participants for a Q-Sort, than proportionality in a random-sampling sense (Addams 2000).

## A Q Method Analysis of Internet Users

The following analysis presents the preliminary interpretation and explanation of commonly shared attitudes or discourses evident in Internet users' willingness, or unwillingness, to participate in social change strategies online. Table 2 shows the correlations between the respondents and the factors. The five factors, or profiles of Internet users are best understood by analysing how the statements that characterise each individual factor fit together.

---

[7] Participants were given three scenarios to which they could relate their involvement and experience of the Internet. Moderate level experience is identified with information exchanges only, an intermediate level included information exchanges, but extended this to include relational exchanges such as maintenance of relationships with family and friends online. Those with a high levels of involvement online included the previous two levels, and also experienced relational exchanges with family and friends and support and collaboration with strangers online. These people also identified with a commitment to online communities of interest.

Particularly important are the statements at the extremes. In this case we have used an opinion continuum ranging *most agree* (+4) to *most disagree* (-4).

In Q method it is the participants performing the Q sorts who are the "variables" being correlated and subsequently factored. Interpretations are based on factor arrays and factor scores rather than loadings, which are typically used in factor analysis based on categorical variables. The final statistical step in the procedure is the analysis of factor arrays (which are essentially composite Q sorts) by calculating factor scores. A factor score is the score gained by each item or statement of the Q set as a kind of weighted average of the scores, given that statement by the Q sorts associated with the factor (Brown 1993: 27). Normalised factor scores may be used to make distinctions between the profiles, however, it is quite adequate to argue in terms of rounded scores (+4 to -4) since as a general rule, differences in scores of 2 or more are considered significant (*p<0.01*) (Addams 2000:32).

*Table 2: Correlation Between Subjects and Factors*

| Internet User | Factor | | | | | Internet User | Factor | | | | |
|---|---|---|---|---|---|---|---|---|---|---|---|
| | A | B | C | D | E | | A | B | C | D | E |
| User01 | 0.701* | 0.123 | 0.096 | -0.099 | 0.350 | User17 | 0.197 | -0.119 | 0.368 | -0.152 | 0.484* |
| User02 | 0.805* | -0.037 | -0.097 | 0.096 | -0.089 | User18 | 0.201 | -0.222 | 0.207 | 0.635* | -0.231 |
| User03 | 0.078 | 0.113 | 0.779* | -0.019 | 0.073 | User19 | 0.222 | -0.140 | 0.112 | 0.649* | 0.331 |
| User04 | 0.684* | -0.227 | 0.291 | 0.045 | -0.025 | User20 | 0.732* | 0.148 | 0.164 | 0.157 | -0.062 |
| User05 | 0.559* | 0.348 | 0.201* | -0.044 | 0.200 | User21 | -0.047 | -0.072 | -0.252 | -0.211 | -0.625* |
| User06 | -0.219 | 0.627* | 0.286 | 0.151 | -0.144 | User22 | 0.773* | -0.085 | -0.003 | -0.057 | 0.041 |
| User07 | 0.740* | 0.002 | -0.052 | 0.344 | 0.191 | User23 | 0.649* | 0.149 | 0.070 | 0.239 | -0.032 |
| User08 | -0.185 | 0.378 | 0.049 | 0.692* | 0.233 | User24 | 0.152 | 0.425 | -0.174 | 0.491* | -0.046 |
| User09 | 0.739* | -0.017 | 0.153 | -0.066 | 0.317 | User25 | 0.332 | 0.273 | 0.219 | 0.449 | 0.211 |
| User10 | 0.457 | 0.673* | 0.154 | 0.078 | -0.051 | User26 | 0.218 | 0.599* | 0.045 | -0.0376 | 0.204 |
| User11 | 0.214 | 0.264 | 0.646* | 0.039 | 0.014 | User27 | -0.245 | 0.564* | -0.404 | -0.013 | -0.096 |
| User11 | 0.214 | 0.264 | 0.646* | 0.039 | 0.014 | User28 | -0.120 | -0.215 | 0.580* | 0.068 | -0.062 |
| User11 | 0.214 | 0.264 | 0.646* | 0.039 | 0.014 | User29 | 0.102 | 0.040 | 0.602* | 0.241 | 0.081 |
| User12 | 0.179 | 0.260 | 0.240 | -0.079 | 0.337 | User30 | 0.493* | 0.380 | -0.228 | 0.205 | -0.022 |
| User13 | 0.553* | -0.019 | 0.422 | -0.336 | 0.070 | User31 | 0.065 | 0.530 | 0.222 | 0.541 | 0.028 |
| User14 | 0.154 | 0.106 | 0.608* | 0.091 | 0.233 | User32 | 0.724* | 0.144 | 0.355 | 0.064 | 0.097 |
| User15 | 0.716* | 0.203 | 0.020 | 0.188 | 0.340 | | | | | | |
| User16 | 0.071 | -0.061 | 0.412 | 0.139 | 0.758* | % expl.Var. | 22 | 9 | 11 | 8 | 7 |

Note: loadings marked with an asterisk (*) are significant.

The statements associated with each profile are discussed in the following section. Where appropriate, statements that distinguish each profile from the other are inserted in the text. The sample number of each statement follows it in brackets. For ease of comparison, the score of all five profiles follow each statement throughout.

### Profile A: The Techno-Optimist

The *Techno-Optimist* profile illustrates the beliefs and attitudes of Internet users who are positive about online exchanges and interactions. Overall this profile is indicative of those Internet users who value the potential for social gathering online, based on the ability to cooperate and be involved in online friendships. Their optimism is based on strong beliefs about communication and cooperation on the Net. They perceive the Net to be "creating communication where there was none" and trust having friends online who can give them assistance. In fact their strong disagreement with *Statement 36* (see Table 3) demonstrates that the Net has lived up to positive expectations. Also, as users of the technology they value the quality and usefulness of information on the Net, contrary to attitudes expressed in *Statement 44*.

*Table 3: Defining Techno-optimist beliefs and attitudes*

| Statement | Profile A | Profile B | Profile C | Profile D | Profile E |
|---|---|---|---|---|---|
| I'm very positive about what the Internet can offer on a communication basis for people … it's creating communication where there was none. (47) | +4 | +1 | 0 | 0 | -4 |
| Having friends online doesn't help at the end of the day because these friends can't really get help from, or give assistance. (40) | -4 | -2 | 0 | -2 | 0 |
| It takes a lot of time to find information on the Internet because a lot of the stuff on the Internet is either wrong, irrelevant or not useful anyway. (44) | -3 | 0 | 3 | 2 | -1 |
| When I actually got to use the Internet it wasn't as wonderful as I thought it was going to be. (36) | -3 | +1 | 0 | 0 | -1 |

To a lesser extent the *Techno-Optimist* values their ability to use the interactive nature of the Internet to inform their views and opinions. They see the Net as having the potential to empower people to have more control over their actions and decision-making and believe Internet users can influence change by becoming involved in social actions, such as online demonstrations (see Table 4). These attitudes enrich the *Techno-Optimist's* profile as they orientate beliefs into online activity.

*Table 4: Secondary Techno-Optimist beliefs and attitudes*

| Statement | Profile A | Profile B | Profile C | Profile D | Profile E |
|---|---|---|---|---|---|
| I think the Internet's good because it's interactive, there is a discussion back and forth and you can actually have discussion on a subject, you're not just being told things, you're discussing them with someone. So that's certainly leads to more in-depth and more rounded views than what you get on the six o'clock news at nights. (20) | +2 | -1 | -2 | -3 | 0 |
| The Internet is empowering people to have more control over what they're seeing, or doing, or thinking about. (50) | +2 | -2 | 0 | -2 | -1 |
| The proof is in the pudding, isn't it actually about getting people to go to the rally, to actually front up. It's all very well sitting at your desk and supporting action, but isn't it more important to actually be there? (27) | -2 | 0 | +1 | 0 | -1 |

### Profile B: The Techno-Realist

The *Techno-Realist* profile highlights some of the ambivalent attitudes users of new communication technologies have towards the Internet. This profile illustrates the fact that some Internet users believe the Net is not necessarily safe or accessible, and that issues around anonymity mediate their attitude towards interpersonal communication online.

The *Techno-Realist* is conscious of the social implications for those who cannot afford access to the Internet (*Statement 21*). On one hand the *Techno-Realist* recognizes the new opportunities the Internet provides for human intervention and doing things together online. For example, they strongly agree with *Statement 6* (Table 5) that people can use a slightly anonymous persona online to talk about deeply, personal information. However, they also feel that the Internet may not offer a safe environment to talk about sensitive issues (*Statement*

*30*). In fact, they believe the Internet doesn't provides a personal space which is any more liberating than other forums where people can talk about sensitive issues (see Table 5).

*Table 5: Primary and Secondary Techno-Realist beliefs and attitudes*

| Statement | Profile A | Profile B | Profile C | Profile D | Profile E |
|---|---|---|---|---|---|
| I'm not particularly optimistic about the Internet in the sense that like any powerful thing, it's about whether or not you can actually afford to access it. (21) | 0 | +4 | -1 | -1 | -1 |
| I think people talking about some deeply, personal, disturbing material is enabled by being able to assume a slightly anonymous persona online. (6) | +1 | +3 | +1 | 0 | 0 |
| I think the Internet's liberating; it lets people talk about issues that they wouldn't talk about in other circumstances. (38) | +1 | -3 | 0 | +2 | 0 |
| You need a safe environment to be able to talk about sensitive issues online. I don't know if that's possible on the Internet. (30) | -2 | +2 | 0 | -2 | 0 |

The picture of the *Techno-Realists* is further elaborated by considering other attitudes significant to their attitude profile (see Table 6). These attitudes are also shared by *Information Gatherers* (Profile C), so are not definitive in describing the *Techo-Realist*, they however enrich our interpretation of the *Techno-Realist* profile. As users of the Internet, *Techno-Realists* do not identify with assuming online identities as they strongly disagreed with the belief that "you can be anybody online" (*Statement 11*). In addition, interpretation of *Statement 19* also highlights that *Techno-Realists* prefer to conduct personal and social relations in face-to-face situations, not via online communication. It is evident that *Techno-Realist* attitudes to anonymity mediate their participation in interpersonal communication online.

*Table 6: Beliefs and attitudes shared by Techno-Realists and Information Gatherers*

| Statement | | Profile A | Profile B | Profile C | Profile D | Profile E |
|---|---|---|---|---|---|---|
| 19 | So much of my personal and social interactions with close friends happen electronically. It's a real frustration for me when I have a friend that's part of my immediate social group that doesn't have email contact. | -1 | -4 | -4 | +2 | +1 |
| 11 | I like being anonymous online, I can be anybody. | -1 | -4 | -3 | 0 | 0 |

### Profile C: The Information Gatherer

The *Information-Gatherer* is the antithesis of those Internet users who gather online to build relationships and create community. In fact, these users see the Internet as information, not the network of people connected through it (see Table 7) and do not find the idea of connecting with like-minded people online at all important (*Statement 15*). The *Information-Gatherers'* lower level of involvement online however is not because of concerns about issues such as privacy. They strongly disagree with the belief that email is not private (*Statement 39*).

*Table 7: Defining Information Gatherer beliefs and attitudes*

| Statement | Profile A | Profile B | Profile C | Profile D | Profile E |
|---|---|---|---|---|---|
| I don't think the Internet is people, I think it's just information. (17) | -3 | -4 | +3 | -1 | -1 |
| The thing I find really exciting about the Internet is the number of people I can network and connect with. It means that you can actually find a group of people that you will be able to have something in common with. And that's what gives me hope for it. (15) | +2 | 0 | -4 | +1 | +2 |
| I would never write anything private in an email, I don't think that they're private. (39) | -2 | -1 | -4 | 0 | +4 |

The *Information-Gatherer* is not interested in developing interpersonal relationships online. Moderate disagreement with *Statement 29* (see Table 8) and strong disagreement with *Statement 19* (see Table 6) in relation to maintenance of personal and social interactions online, supports this interpretation. The *Information-Gatherer* is optimistic about the Internet. We argue the case for constructing the *Information-Gatherer* not as a passive consumer of information, but rather an active, information-seeker, searching for online information and knowledge. The point is that the *Information-Gatherer* is "just carving off the slice" they want.

*Table 8: Secondary Information Gather beliefs and attitudes*

| Statement | Profile A | Profile B | Profile C | Profile D | Profile E |
|---|---|---|---|---|---|
| I'm optimistic about the Internet because I see a lot of potential out of it, but I see that you've got not a lot of choice but to be enthusiastic and just carve off your slice of it. (1) | 0 | 0 | +2 | 0 | 0 |
| By using the Internet I keep in touch with people more than I ever would have done before. (29) | +1 | +3 | -2 | +4 | +1 |

## *Profile D: The Anti-Political Opportunist*

The Internet has been credited with changing the dynamics of popular contention (Ayres 1999). Traditionally political protests have relied heavily on people rallying in the streets to support a cause. However the "Internet is altering this dynamic by electronically promoting the diffusion of protest ideas and tactics efficiently and quickly across the globe" (Ayres 1999: 132). This is not a sentiment that is shared by the *Anti-Political Opportunist*. These Internet users actively seek information about issues that are important to them (see Table 9), but they have negative attitudes towards online protest activities. The *Anti-Political Opportunist* strongly disagrees with the belief that the Internet politicises people through participation in the exchange of information and dialogue. They also believe that the circulation of email protests are an ineffective protest tactic (*Statement 25*).

*Table 9: Defining Anti-Political Opportunist beliefs and attitudes*

| Statement | Profile A | Profile B | Profile C | Profile D | Profile E |
|---|---|---|---|---|---|
| I use the Net to find out as much as I can about issues that are important to me. (4) | +1 | 0 | 0 | +4 | +2 |
| I don't tend to sign online petitions because they don't do a lot of use. (25) | -1 | 0 | -1 | +3 | 0 |
| I guess the Internet does politicise people, because they're taking the time online to talk about issues and the big picture. (49) | +1 | 0 | -1 | -4 | 0 |

## Profile E: The Techno-Sceptics

To some extent the Internet has created a consciousness that the world we live in today is highly risky and that the pervasive use of the Internet in our daily lives has increased uncertainty (Slevin 2000). This perspective of the Internet's role in society is illustrated by the *Techno-Sceptic* profile. For example, the *Techno-Sceptic* has strong concerns about the privacy of email exchanges (*Statement 39*) and believes that individual privacy is more at risk online, than offline (*Statement 31*). Their lower levels of trust are demonstrated through strong agreement with the belief that people will drive a cause online, not because they are committed to the issue, but because they can be anonymous (see Table 10). Trust and privacy concerns mediate the *Techno-Sceptic's* involvement in online exchanges.

*Table 10: Defining Techno-Sceptics attitudes and beliefs*

| Statement | Profile A | Profile B | Profile C | Profile D | Profile E |
|---|---|---|---|---|---|
| I would never write anything private in an email, I don't think that they're private. (39) | -2 | -1 | -4 | 0 | +4 |
| People could legitimately drive a cause online that they were not interested in, you know, from a totally anonymous point of view. (10) | 0 | -2 | 0 | 0 | +3 |
| I'm very positive about what the Internet can offer on a communication basis for people … it's creating communication where there was none. (47) | +4 | +1 | 0 | 0 | -4 |
| People have access to your name, address, the things you like, your movements, where you go, what you do. But these risks exist anyway. (31) | 0 | +1 | +3 | +2 | -3 |

Consideration of other beliefs and values enrich our understanding of the *Techno-Sceptic* profile. Disagreement with *Statement 34* (Table 11) could be illustrative of an underlying determinist view of the Internet. We would argue that these users see the technology as controlling their interaction, rather than seeing the Internet as conditional and flexible, and as a technology that could contribute to community and personal empowerment. Lower levels of trust also mediate these beliefs. It is interesting to note that the *Techno-Sceptic* has a slightly positive attitude towards gathering information from online discussions, which contradicts their beliefs about the Internet creating more communication. Are the *Techno-Sceptics* the lurkers in online communities?

*Table 11: Secondary Techno-Sceptic beliefs and attitudes*

| Statement | Profile A | Profile B | Profile C | Profile D | Profile E |
|---|---|---|---|---|---|
| How good the Internet will be depends on how a person uses it and how the person actually uses it to better themselves for the good of society and things like that. (34) | +3 | +4 | +2 | 0 | -2 |
| I find online discussions very much like small talk, even when we were talking about stuff that was meaningful for people and interesting. I didn't find that personally the way I liked to gather information. (5) | 0 | 0 | +1 | 0 | -2 |

## Discussion

The development of online communication strategies requires both "transactional" elements and "relational" elements (Armstrong and Hagel 1996). We argue that online campaign strategies that prioritise the importance of the social functions of the Internet, relative to the informational functions (Sproull and Faraj 1997) will be more successful. As a result, online campaigns will be more interactive rather than passive, and thus create more involving exchanges between social changes agencies and their target adopters and markets.

Learnings from the sociology literature support our position on the value of increasing interactivity and relational exchanges in online social cause campaigns. These learnings were supported by findings from interview and focus group data. We found that Internet users interviewed demonstrated specific attitude and belief structures that profiled their individual proneness to participate actively in online exchanges. Thus their involvement and participation in online exchanges dealing with either sensitive issues, or social causes, are based on individual perceptions of risk and trust concerning Internet interactions and their individual personal situation and other factors, such as attitudes and beliefs towards interactive technologies in general, along with a range of intra-personal, social and economic motives.

To explicate the underlying belief and attitude structures of Internet users we have developed five idealised profiles using a Q method. As a research approach, Q method has not been used widely in the marketing field. However, we have found it to be a highly useful technique which has enabled us to engage the social discourses around peoples' beliefs and attitudes towards online interaction. Our analysis thus discusses more directly the social factors that shape Internet-based exchanges. We can do this because the statements selected from the interviews and focus groups situate actions and interpretations that people give to the Internet (attitudes), and other statements outline how people conceptualise, or make sense of their online interaction and sense of community, maintenance of online relationships, bonds of trust and reciprocity.

### *A sample of professionals who use the Internet*

The Q-technique that ranked the five profiles emanated from research participants' own subjective perspectives of the Internet as a tool for social change strategies. Thus, each profile naturally categorises the thought and sentiment (Brown et all 1999) within the group of educated, professional respondents. We make these statements to simply acknowledge that the factors are a function of the lived experience of these professionals. They are also purely inductive in the sense that their number and character have been induced from those individuals who produced them (Brown et all 1999).

Some researchers would argue that the research sample we have used in this research is limited in its power to describe the broader population of Internet users. We partially agree with this assessment, and in future studies will select a larger number of Internet users with higher levels of involvement in relational exchanges. However, it is important to note here that the use of Q method is not concerned with representing all of the discursive constructors of involvement and participation in the Internet and online social change agendas.

The findings from this pilot study present an interesting insight into the segment of Internet users with professional backgrounds, who have higher levels of education and income. It is interesting to find that the models generated represent underlying information- seeking behaviour, and an absence of networking behaviour. Some of the profiles demonstrate bonding (Putnam 2000) relationships, which focus on maintaining relationships with family and friends using the Internet. However, none of profiles are demonstrative of the bridging and linking relationships, which Putnam (2000; 2001) argues are critical to the future civic and physical health of our society(ies). Bridging and linking networks are important as they engage in developing relationships across diverse groups of society stakeholders, which are critical in effecting change and build trust in our communities.

### *Future online campaign strategies*

Many social scientists have reached the pessimistic conclusion that information campaigns are ineffective and often fail to influence the public's knowledge about or attitudes towards important social and health behaviours. Yet the significance of information-seeking behaviour evident in the profile of Internet users we discussed throughout this paper should encourage social marketers to embrace information richness in their social change strategies. This is especially if they are developing and implementing strategies that target segments of the professional population who regularly access the Internet to gather information on issues that are important to them.

## Conclusion

According to Ackermann (cited in Brown 1999: 169), what is required for progress in the human sciences is "not simply more data … , as many empiricists have stated, but new instrumentation for obtaining data … so that more exhaustive explanatory possibilities can be tried." By using Q methodology we have attempted to pursue this sentiment in profiling current Internet users' beliefs and attitudes towards online social change strategies. Findings from the research are insightful about Internet users from two perspectives. Firstly, the research provides an interesting insight into the attitudes of professionals towards using the Internet to facilitate their participation in social change strategies. Secondly, the findings contribute to the broader research profile of Internet users. The five profiles explicated in our discussion extend the understanding of Internet users' transactional and relational behaviour online, beyond a purely commercial experience, to include important social experiences. In doing so we hope to have stimulated some new ideas and debates concerning Internet users' willingness to participate in social change strategies in online environments.